# Fully dense MgB$_2$ superconductor textured by hot deformation

A. Handstein*, D. Hinz, G. Fuchs, K.-H. Müller, K. Nenkov[1], O. Gutfleisch, V.N. Narozhnyi[2], L. Schultz

*Institut für Festkörper- und Werkstoffforschung Dresden, Postfach 270116, D-01171 Dresden, Germany*

* e-mail address: a.handstein@ifw-dresden.de

[1] On leave from International Laboratory of High Magnetic Fields Wroclaw, Poland; ISSP-BAS, Sofia, Bulgaria

[2] On leave from Institute for High Pressure Physics, Russian Academy of Sciences, Troitsk, Russia

**Abstract**

Bulk textured MgB$_2$ material of nearly full density showing a weak c-axis alignment of the hexagonal MgB$_2$ grains parallel to the pressure direction was obtained by hot deformation of a stoichiometric MgB$_2$ pellet prepared by a gas-solid reaction. The texture of the material was verified by comparing the x-ray diffraction patterns of the hot deformed material with isotropic MgB$_2$ powder. A small, but distinct anisotropy of the upper critical field up to $H_{c2}^{a,b}/H_{c2}^{c} \sim 1.2$ depending on degree of texture was found by resistance and susceptibility measurements. No anisotropy of the critical current density determined from magnetization measurements was found for the textured material.

**1. Introduction**

The recent discovery of superconductivity in MgB$_2$ [1] at temperatures as high as 40 K has stimulated considerable interest in this system. MgB$_2$ powder with good superconducting properties has been reported [2]. Great efforts were made to produce fully dense MgB$_2$ material. MgB$_2$ wires with a density of $\rho = 2.4$ g/cm³ were produced by a diffusion process of Mg into B fibres [3] which compares with a theoretical density of about 2.6 g/cm³. Sintering of MgB$_2$ powder at 1250°C and a high pressure of 3.5 GPa led to material reaching the theoretical density [4]. The upper critical field anisotropy of MgB$_2$ was investigated on samples of aligned MgB$_2$ crystallites [5] and MgB$_2$ single crystals [6].

MgB$_2$ crystallizes in the hexagonal AlB$_2$ type structure (space group P6/mmm) with unit cell parameters $a = 0.308136$ nm and $c = 0.351782$ nm [7]. This anisotropic structure has given the motivation to investigate the formation of texture by hot deformation as known from Nd-Fe-B materials (e.g. [8]). In the present study, the properties of MgB$_2$ prepared by the conventional gas-solid reaction and subsequently deformed by a die-upsetting process have been investigated.

## 2. Experimental

For preparing bulk MgB$_2$ pellets, a stoichiometric mixture of crystalline B powder and Mg was pressed into cylindrical pellets with a diameter of 8 mm. The pellets were wrapped in Ta foil and sealed in a quartz ampoule under an Ar gas pressure of 180 mbar. The ampoule was heated stepwise from 750°C to 950°C during one hour and then annealed at 950°C for 2 h. After this gas-solid reaction the MgB$_2$ pellets enlarged in their diameter to about 8.5 mm were hot deformed by die-upsetting in an Ar atmosphere at temperatures from 850 to 900°C resulting in a height reduction of about 70%. The chosen strain rate of $1 \times 10^{-3} s^{-1}$ could not be realized permanently during the die-upset process because the deformation proceeded discontinuously. The maximum stress was 450 MPa. Fig. 1 shows the applied force and the length reduction of the sample in dependence on time demonstrating the inhomogeneous material flow during the deformation. It is known that the MgB$_2$ material becomes very porous after the gas-solid reaction, which can be explained by the evaporation of Mg particles. It is obvious that during the stepwise deformation of the material under the influence of the applied force these porosities disappear. It is interesting to note that the steps in the force vs. time and in the corresponding stress vs. strain curves reflecting the discontinuous deformation of the material (see Fig. 1) vanish with prolonged deformation. It can be assumed that only here a continuous stress-strain-regime is reached and effective texturing takes place.

The samples were characterized by susceptibility and resistance measurements that were performed in an ac susceptometer type 7229 (Lake Shore). X-ray diffraction measurements were made with a Bragg-Brentano diffractometer using Cu K$_\alpha$ radiation. Optical microscope observation was carried out on polished sample planes perpendicular to the pressure direction and, after cross cutting the flat samples, parallel to the pressure direction.

## 3. Results and discussion

Optical investigation revealed a very different microstructure in dependence on the observed planes. The observation on the imaging plane perpendicular to the pressure direction shows large grains after die-upsetting at 850°C (Fig. 2a), whereas only smaller grains are visible in the cut plane (i.e. parallel to the pressure direction, see Fig. 2b). Obviously, grains of platelet shape were oriented parallel to the directly pressed sample plane due to die-upsetting. This result was elucidated by x-ray diffraction (XRD) investigation. The XRD pattern of a sample deformed at 900°C (Fig. 3b) is consistent with that of $MgB_2$ powder (Fig. 3a). The minor reflections can be identified as pure Mg and MgO. The comparison of the intensity of essential peaks with those of the isotropic powder indicates a weak alignment of the *c*-axis of the hexagonal $MgB_2$ grains along the direction of pressure indicated by an enhancement of the (00l) intensities and a decrease of the (hk0) reflections. That means an alignment of the *c* axis of some grains parallel to the pressure direction was induced due to hot deformation (in the following the pressure direction is named as the c axis of textured $MgB_2$ material).

An appropriate characterization of the texture has been proposed by Lotgering [9] who defined an alignment degree A by *A = (P-P₀)/(1-P₀)* with *P = ΣI₍₀₀l₎/ΣI₍hkl₎* for the textured sample (*I* is the peak intensity of a Bragg reflection) and the corresponding expression *P₀* for the isotropic sample. For further simplification, in this study only two reflections, for instance to (110) and (002), will be considered using the alternative alignment degree

$$\tilde{A} = \frac{\frac{1}{1+I_{(110)}/I_{(002)}} - \frac{1}{1+I_{0\,(110)}/I_{0\,(002)}}}{1 - \frac{1}{1+I_{0\,(110)}/I_{0\,(002)}}} \qquad (1)$$

Ideal alignment leads to $\tilde{A} = 1$ whereas a random orientation results in $\tilde{A} = 0$. In Table 1 the results of two samples with different degrees of texture are compared. The value of $\tilde{A} = 0.49$ obtained from the intensity ratios of the (110) and (002) reflections of sample I indicates that the texture in this sample is more pronounced than in sample II for which a lower value of 0.23 is obtained. This behaviour is also confirmed for the (100) and (002) reflections as shown in Table 1. It is assumed that the formation of texture is strongly influenced by temperature and degree of deformation, density of the starting material and other processing parameters. A more detailed analysis of texture is in progress.

The resistivities of the investigated samples (c.f. Table 1) were found to be reduced from 29.8 µΩcm (sample I) and 27.5 µΩcm (sample II) at 300 K to about 9 µΩcm (sample I) and 6 µΩcm (sample II) at 40 K resulting in residual resistance ratios (*RRR*) of approximately 3.4 (sample I) and 4.6 (sample II). For the superconducting transition temperature, values of $T_c = 38.5$ K (sample I) and $T_c = 39.2$ K (sample II) were determined from *ac* susceptibility data using the onset temperature of the superconducting transition. Similar results were obtained from resistance measurements. In Fig. 4, the temperature dependence of the resistance of sample I is shown for several magnetic fields up to 7.5 T applied parallel and perpendicular to the *c*-axis of the textured sample. A distinct anisotropy develops for applied magnetic fields $\mu_o H \geq 1$ T that increases with increasing field. Fig. 5 shows upper critical field data determined at 10% of the resistance in the normal state. The comparison of the $H_{c2}(T)$ curves in Fig. 5 (results of sample II, c.f. Table1) shows clearly that the upper critical field is enhanced when the magnetic field is applied parallel to the *ab* plane. This is in agreement with results for aligned MgB$_2$ crystallites [5] and MgB$_2$ single crystals [6] for which the ratio $H_{c2}^{a,b} / H_{c2}^{c}$ was found to be $H_{c2}^{a,b} / H_{c2}^{c} \sim 1.7$ and $H_{c2}^{a,b} / H_{c2}^{c} \sim 2.6$, respectively. For our textured MgB$_2$ material, this ratio is $H_{c2}^{a,b} / H_{c2}^{c} \sim 1.2$ for sample I with a higher degree of texture and $H_{c2}^{a,b} / H_{c2}^{c} \sim 1.1$ for sample II (see Table I). Although the anisotropy in MgB$_2$ is much smaller than in the highly anisotropic high-$T_c$ cuprates, it is considerably larger than in the quaternary borocarbides. For instance, a ratio of $H_{c2}^{a,b}/H_{c2}^{c} \sim 1.13$ has been found for single-crystalline LuNi$_2$B$_2$C [10].

A peculiarity of the $H_{c2}(T)$ curves shown in Fig. 5 is their pronounced positive curvature near $T_c$. Similar results have been already observed for non-textured MgB$_2$ samples [11, 12]. Such a positive curvature of $H_{c2}(T)$ near $T_c$ is a typical feature observed for the non-magnetic rare-earth nickel borocarbides $R$Ni$_2$B$_2$C ($R$=Y, Lu) and can be explained by taking into account the dispersion of the Fermi velocity using an effective two-band model for superconductors in the clean limit [13]. This model can be successfully applied to MgB$_2$, as was shown recently [14]. We conclude that also our MgB$_2$ samples are within the clean limit in spite of the rather moderate RRR value of 4.6. It is interesting to note that a considerable reduction of the positive curvature of $H_{c2}(T)$ was observed for radiation-disordered MgB$_2$ samples [15].

In contrast to the upper critical field, no anisotropy was found for the critical current density of the textured samples which was determined form magnetization measurements. The relation $j_c = 30\ \Delta M/l$ for a square plate with lateral dimensions $l$ in a perpendicular magnetic field was used to calculate the critical current density $j_c$ (in A/cm$^2$) from the magnetization loop width $\Delta M$ (in emu/cm$^3$) and the length $l$ (in cm). The field dependence of the critical current density of sample II is shown in Fig. 6 for several temperatures. The critical current density decreases in a wide field range almost exponentially with increasing magnetic field which is in agreement with data reported by Dhallé et al. [16] for MgB$_2$ samples with a mean grain size of 6.5 µm. The slightly lower $j_c$ values for our textured MgB$_2$ sample compared to these samples [16] might be connected with the relatively coarse grained structure of our material (see Fig. 2) and the reduced number of grain boundaries which probably act as pinning centers in MgB$_2$.

## 3. Summary

Bulk $MgB_2$ material of nearly full density was prepared from porous $MgB_2$ pellets by die-upsetting in an Ar atmosphere at temperatures from 850 to 900°C. A weak *c*-axis alignment along the pressure direction was found in the hot deformed samples as confirmed by x-ray diffraction as well by measuring the anisotropy of the upper critical field $H_{c2}(T)$. A ratio of $H_{c2}^{a,b}/H_{c2}^{c} \sim 1.2$ was found for the textured $MgB_2$ material which is expected to increase with increasing deformation degree and temperature. It is assumed that the slightly lower $j_c$ values of the textured $MgB_2$ material, compared with the higher values obtained from finely grained material [16], are caused by the relatively coarse grained structure and the reduced number of grain boundaries which probably act as pinning centers in this textured $MgB_2$.

**Acknowledgement**

The authors thank S.V. Shulga and S.-L. Drechsler for valuable discussions. They are also grateful to Mrs. M. Gründlich and B. Opitz for making the optical microstructure and x-ray investigations, respectively.

**References**

[1] J. Nagamatsu, N. Nakagawa, T. Muranaka, Y. Zenitani, Nature **410** (2001) 63.

[2] S.L. Bud´ko, G. Lapertot, C. Petrovic, C.E. Cuningham, N. Anderson, P.C. Canfield, Phys. Rev. Letters **86** (2001) 1877.

[3] P.C. Canfield, D.K. Finnemore, S.L. Bud´ko, J.E. Ostenson, G. Lapertot, C.E. Cunningham, C. Petrovic, Phys. Rev. Letters **86** (2001) 2423.

[4] Y. Takano H. Takeya, H. Fujii, H. Kumakura, T. Hatano, K. Togano, cond-mat/0102167, 2001.

[5] O.F. de Lima, R.A. Ribeiro, M.A. Avila, C.A. Cardoso, A.A. Coelho, cond-mat/0103287, 2001.

[6] M Xu, H. Kitazawa, Y. Takano, J.Ye, K. Nishida, H. Abe, A. Matsushita, and G. Kido, cond-mat/0105271.

[7] L.-H. He, G.-Q. Hu, Chinese Physics **10** (2001) 343.

[8] R.W. Lee, Appl. Phys. Rev. **46** (1985) 790.

[9] F.K. Lotgering, J. Inorg. Nucl. Chem. **9** (1959) 113.

[10] K.D.D. Rathnayaka, A.K. Bahtnagar, A. Parasiris, D.G. Naugle, P.C. Canfield, and B.K. Cho, Phys. Rev. B **55** (1997) 8506.

[11] K.-H. Müller, G. Fuchs, A. Handstein, K. Nenkov, V.N. Narozhnyi, D. Eckert, J. Alloys Comp. **322** (2001) L10.

[12] G. Fuchs, K.-H. Müller, A. Handstein, K. Nenkov, V.N. Narozhnyi, D. Eckert, M. Wolf, L. Schultz, accepted by Sol. St. Comm., cond-mat/0104087.

[13] S.V. Shulga, S.-L. Drechsler, G. Fuchs, K.-H. Müller, K. Winzer, M. Heinecke, K. Krug, Phys. Rev. Lett. **80** (1998) 1730.

[14] S.V. Shulga, S.-L. Drechsler, H. Eschrig, H. Rosner, W. Pickett, cond-mat/0103154, 2001.

[15] A.E. Karkin, V.I. Voronin, T.V. Dyachkova, A.P. Tyutunnik, V.G. Zubkov, Yu.G. Zainulin, and B.N. Goshitskii, cond-mat/0103344, 2001.

[16] M. Dhallé, P. Toulemonde, C. Beneduce, N. Muolino, M. Decroux and R. Flükiger, cond-mat/0104395.

**Figure captions**

Fig. 1 Applied force and length reduction of a MgB$_2$ sample in dependence on time during die-upsetting at 900°C.

Fig. 2. Microstructure of MgB$_2$ after hot deformation at 850°C: imaging plane (a) perpendicular and (b) parallel to the pressure direction.

Fig. 3 Comparison of the x-ray diffraction patterns of (a) MgB$_2$ powder and (b) a MgB$_2$ sample after die-upsetting at 900°C. Minor reflections marked as pure Mg (**x**) and MgO (o).

Fig. 4 Temperature dependence of resistance for several magnetic fields applied parallel (dotted line) and perpendicular (solid line) to the *c*-axis of textured MgB$_2$ (sample I) deformed at 900°C.

Fig. 5 Temperature dependence of the upper critical field $H_{c2}$ of textured MgB$_2$ (sample I) determined from resistance measurements (see Fig. 4; ● - $H // c$, ○ - $H \perp c$). Lines are guides for the eye.

Fig. 6 Field dependence of the critical current density of textured MgB$_2$ (sample II) for magnetic fields $H // c$ at $T = 20$ K, 25 K and 30 K.

**Table**

Table 1: Temperature of hot deformation and comparison of properties of two samples with different degree of texture: texture parameter $\tilde{A}$ (see eq. (1)), residual resistance ratio R$_{40K}$/R$_{300K}$ - *RRR*, superconducting transition temperature $T_c$ and ratio of upper critical fields $H_{c2}^{a,b} / H_{c2}^{c}$).

| Sample | Temperature of hot deformation °C | Alignment degree $\tilde{A}$ determined from | | RRR | $T_c$ K | $H_{c2}^{a,b} / H_{c2}^{c}$ |
|---|---|---|---|---|---|---|
| | | $I_{(110)}/I_{(002)}$ | $I_{(100)}/I_{(002)}$ | | | |
| I | 900 | 0.49 | 0.37 | 3.5 | 38.5 | 1.2 |
| II | 850 | 0.23 | 0.16 | 4.6 | 39.2 | 1.1 |

Table 1

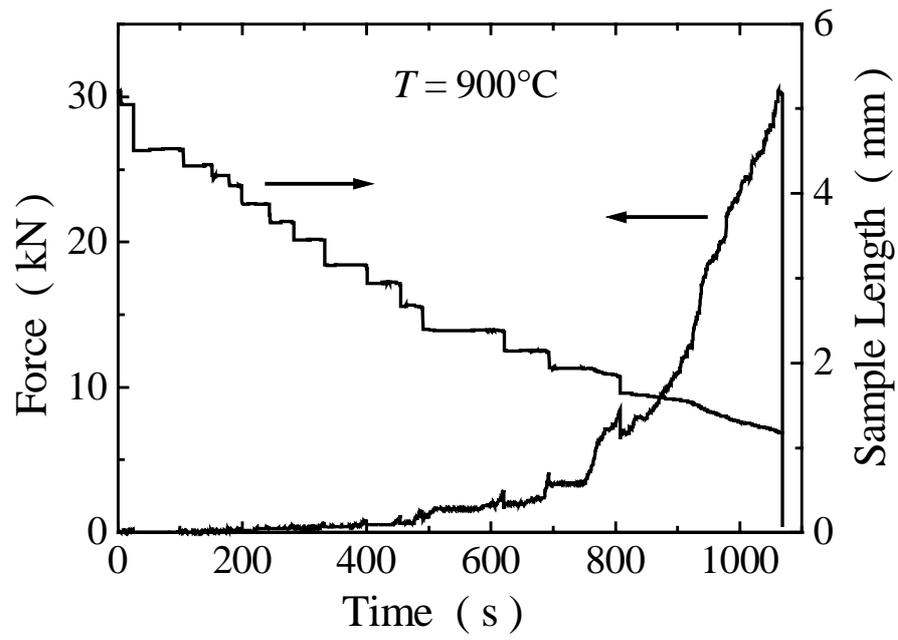

Fig. 1

A. Handstein et al., J. Alloys Comp.

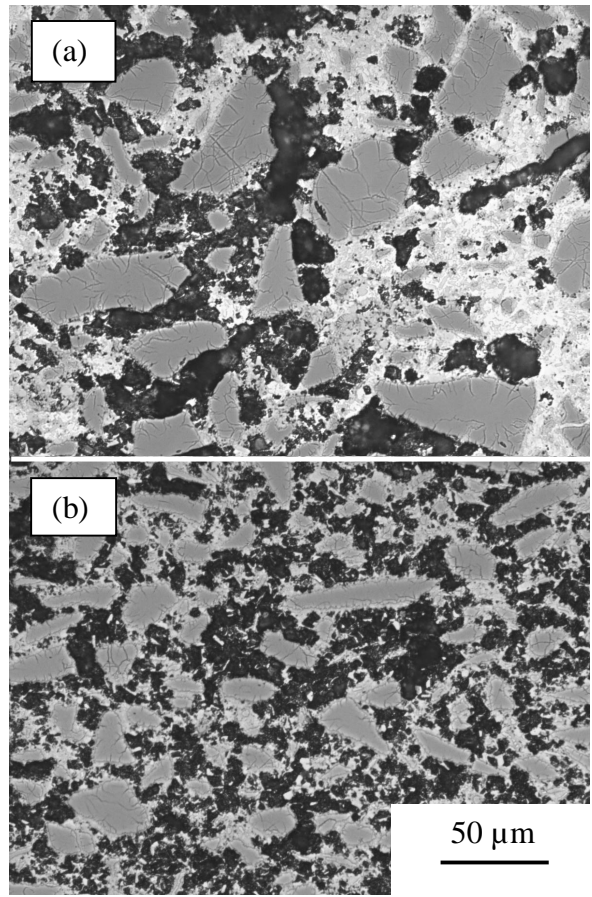

Fig. 2

A. Handstein et al., J. Alloys Comp.

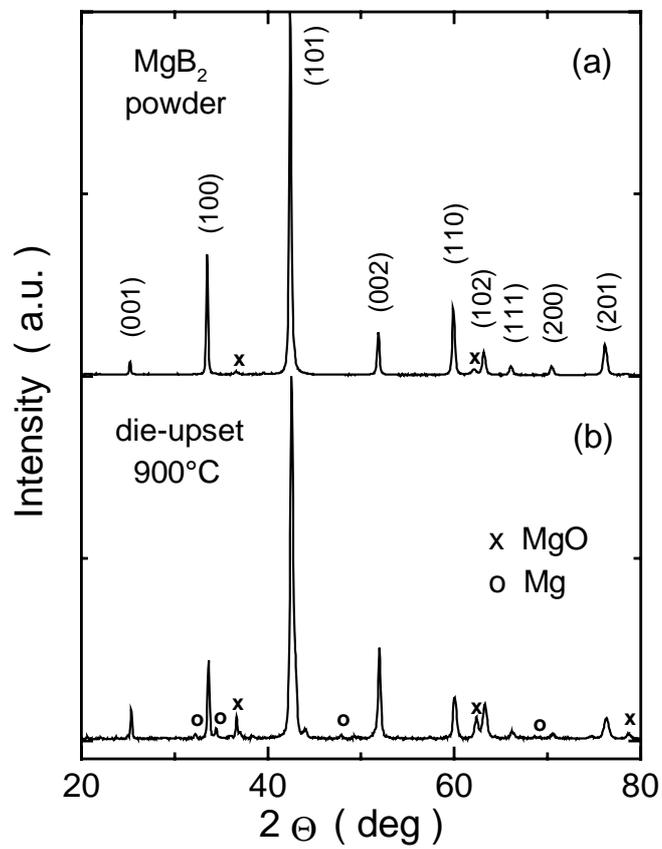

Fig. 3

A. Handstein et al., J. Alloys Comp.

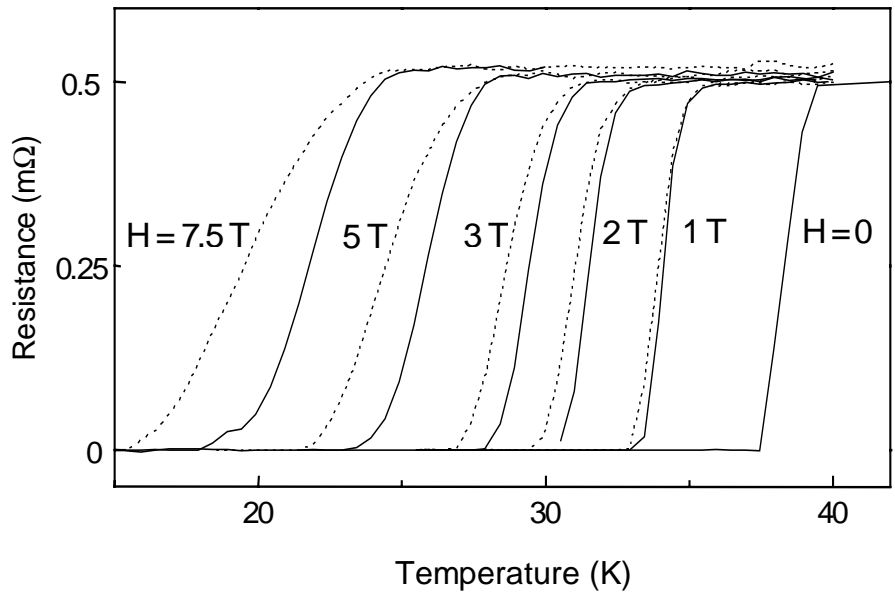

Fig. 4

A. Handstein et al., J. Alloys Comp.

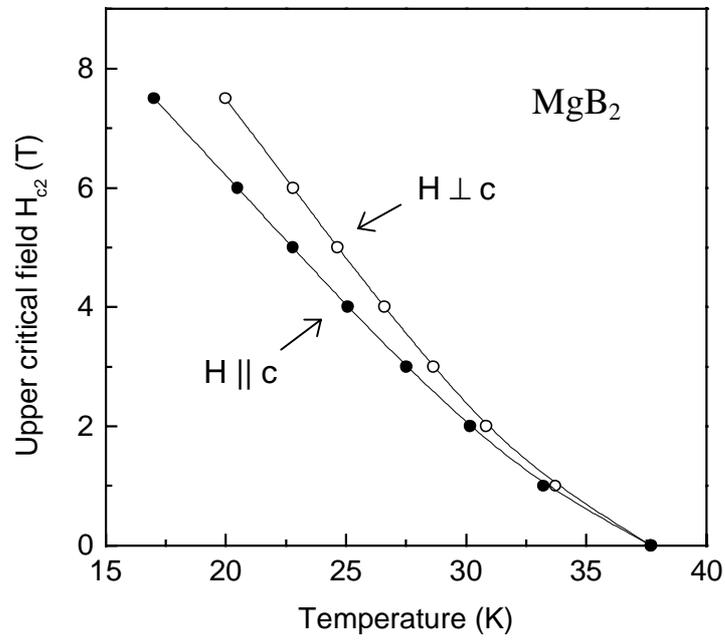

Fig. 5

A. Handstein et al., J. Alloys Comp.

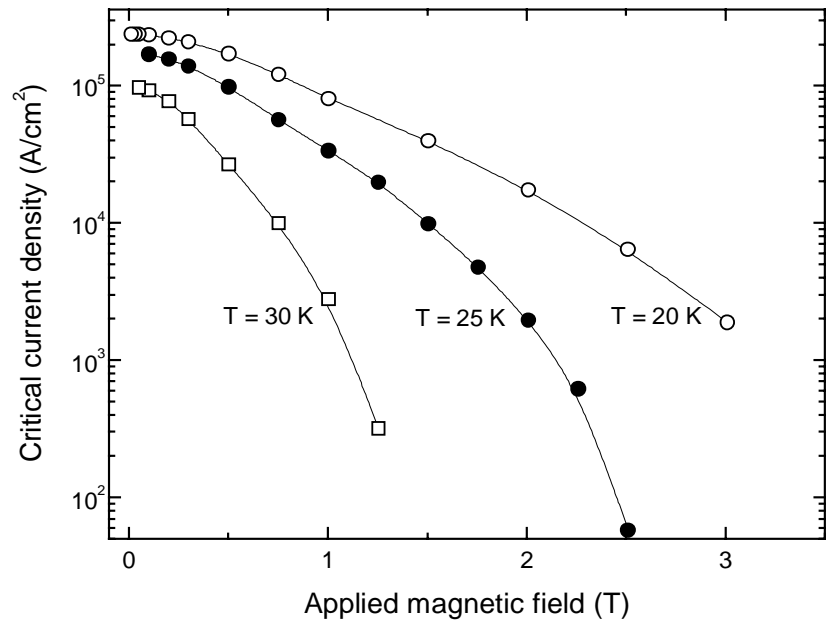

Fig. 6

A. Handstein et al., J. Alloys Comp.